\documentclass[aps,prl,groupedaddress]{revtex4}
\usepackage{latexsym}
\usepackage[dvips]{graphicx}

\newcommand{\eq}{\begin{equation}}
\newcommand{\feq}{\end{equation}}
\newcommand{\eqn}{\begin{eqnarray}}
\newcommand{\feqn}{\end{eqnarray}}
\newcommand{\arr}{\begin{eqnarray*}}
\newcommand{\farr}{\end{eqnarray*}}
\newcommand{\beq}{\begin{equation}}
\newcommand{\eeq}{\end{equation}}
\newcommand{\bea}{\begin{eqnarray}}
\newcommand{\eea}{\end{eqnarray}}

\def\beq{\begin{equation}}
\def\eeq{\end{equation}}
\def\feq{\end{equation}}
\def\bea{\begin{eqnarray}}
\def\eea{\end{eqnarray}}
\def\bc{\begin{displaymath}}
\def\ec{\end{displaymath}}
\def\lb{\label}

\def\la{\lambda}

\def\lb{\label}

\begin{document}
\title{Acoustic horizons for axially  and spherically symmetric fluid flow}

\author{Mariano Cadoni}
\email{mariano.cadoni@ca.infn.it}

\affiliation{Dipartimento di Fisica,
Universit\`a di Cagliari, and INFN, Sezione di Cagliari, Cittadella
Universitaria 09042 Monserrato, Italy}
\author {Paolo Pani}
\email{paolo.pani@ca.infn.it }
\affiliation{Dipartimento di Fisica,
Universit\`a di Cagliari, Cittadella
Universitaria 09042 Monserrato, Italy}


\begin{abstract}
 We  investigate the formation of acoustic horizons  
for an inviscid fluid moving in a pipe in the case of stationary 
and axi-symmetric flow. 
We  show that, differently from what is generally believed, the
acoustic horizon  forms in correspondence of either   a local minimum or  maximum 
of the
flux tube cross-section. Similarly, the external potential is
required to have either a maximum or a minimum at the horizon,
so that the external force has to vanish there.
Choosing  a power-law equation  of state for
the fluid, $P\propto \rho^{n}$, we  solve the equations of the fluid dynamics and
show that the two possibilities are realized
respectively for $n>-1$ and $n<-1$.
These results are  extended also to the case of  spherically symmetric 
flow.
\end{abstract}


\maketitle


\maketitle

Sonic horizons can be generated by the motion of a classical 
fluid \cite{Unruh:1980cg}.
The simplest case is represented by a fluid with space-dependent
velocity flowing trough a tube with  variable section (Laval 
nozzle) \cite{liepmann}.
If the fluid flows in the direction where the tube becomes
narrower,  the fluid velocity  $v$ will increase downstream.  Eventually,
a point will be reached where $v$ is equal to
the local speed of sound $c$ and $v>c$ beyond this point.  An
acoustic perturbation generated in the supersonic region cannot reach
the upstream subsonic region. We have generated a sonic horizon, the acoustic
analogue of a  spacetime horizon. The downstream subsonic region
behaves, as long as only acoustic disturbances are concerned, as the
acoustic analogue of a black hole, a {\it dumb hole}.

Starting from the original proposal of Unruh \cite{Unruh:1980cg},
acoustic (and other condensed matter analogues) black holes
have been used to describe  various
kinematical aspects of general relativity such as event and cosmological
horizons, field theory in curved spacetime, Hawking radiation etc.
\cite{Visser:1993ub,Visser:1997ux,Novello:2002qg,Visser:2001fe,Cardoso:2005ij,
Barcelo:2005fc, Volovik:2000ua, Barcelo:2001ca,fedichev0303,fedichev0304,
Barcelo:2003et,Volovik:2003ga, Barcelo:2004wz, 
Berti:2004ju,Cardoso:2004fi,Nakano:2004ha,Volovik:2003fe}.
Recently,  the fluid/gravity analogy as been extended also at a
dynamical level. It has been used   to gain information about  Einstein's
equations
governing the dynamics of  a spherically symmetric gravitational
black hole \cite{Cadoni:2004my,Cadoni:2005jg}. The lack of direct 
experimental tests
has always been an obstacle for  the research on black holes.
Investigations on ``artificial''
black holes could represent a useful bridge to make black holes
experimentally more accessible.

Although conceptually very simple,  the realization of acoustic
horizons requires fine-tuning of the external potential acting on the
fluid,  of the tube profile and of the initial fluid velocity
\cite{Liberati:2000pt, Visser:2001fe,Cardoso:2005ij}. 
In the case of the Laval nozzle, it 
is widely believed that
the horizon must form exactly  at the  narrowest part of the nozzle
\cite{Visser:2001fe,Sakagami:2001ph, Volovik:2003fe,Cardoso:2005ij}. 
However,  it is not easy  to find necessary
and sufficient conditions for the formation of the horizon. In the 
general case,  the
equations governing the dynamics of the fluid are very difficult to
solve. 

In this paper we investigate the formation of acoustic horizons
in the case of a inviscid fluid with stationary, axially or spherically
symmetric flow.
When considering axi-symmetric fluid motion we will discuss separately
the case of a flux tube of varying
section and the presence of external forces.
We  show that, differently from what is generally believed, the
acoustic horizon  forms in correspondence of either   a local minimum or 
maximum of the
flux tube cross-section. Similarly, the external potential is
required to have either a maximum or a minimum at the horizon,
so that the external force has to vanish there.
Choosing  a power-law equation  of state for
the fluid, $P\propto\rho^{n}$, we solve the equations of the dynamics and
show that the two possibilities are realized
respectively for $n>-1$ and $n<-1$. Moreover, we will determine the
range in the initial conditions for which the horizon effectively
forms.
\vskip 3truemm
\leftline{\bf Axi-symmetric flow}
\vskip 3truemm
Let us first consider  axially symmetric, stationary, inviscid fluid
motion, which is constrained on a pipe.
Indicating with $x$ the coordinate along the symmetry axis,  the fluid
is described by its velocity $v(x)$,   density   $\rho(x)$ and
pressure $P(x)$.
The flux tube is characterized by its cross-section
$A(x)$ and we will assume that external forces, characterized by the
potential $\psi(x))$, act on the fluid. Because of the  symmetries of
the problem  all  the parameters are functions of the coordinate $x$
only. The equations governing the dynamics of the fluid are the
continuity and Euler equations
\beq\lb{de}
\frac{d(\rho A v)}{dx}=0,\quad\quad \rho v
\frac{dv}{dx}=-\frac{dP}{dx}-\rho\frac{d\psi}{dx}.
\feq
We will restrict ourself to the case of isentropic fluid flow
$P=P(\rho)$, which gives for the local speed of sound
$c=\sqrt{\frac{dP}{d\rho}}$.
We will discuss separately the case of a flux tube of variable
cross-section and homogeneous external potential and that of a flux tube of
constant section and variable external potential.
\vskip 3truemm
\leftline{\it{Flux tube with  variable cross-section}}
\vskip 3truemm
Using Eqs. (\ref{de}) with $\psi=const.$, one easily obtain the well-known nozzle
equations \cite{Visser:2001fe,Sakagami:2001ph}
\beq\lb{nozzle}
\frac{1}{A}\frac{dA}{dx}= \frac{1}{v}\frac{dv}{dx}(M^{2}-1),
\feq
where $M=v/c$ is the Mach number.
The nozzle equation (\ref{nozzle}) can be used to discuss the
conditions for the formation of sonic horizons,  i.e of a surface
with $M=1$ separating a subsonic ($M<1$) from a supersonic ($M>1$)
region.
Assuming that $v,v'(x)=dv/dx, A$ are everywhere finite and nonvanishing,
one can easily see from Eq. (\ref{nozzle}) that, the if a horizon
forms,  this necessarily happens in correspondence of a
local extremum of the tube cross-section $A$. Conversely, if the
cross-section $A(x)$ has a local extremum at a point $x_{h}$
necessarily a sonic horizon must form at this point.
Depending on the sign of $v'(x)$ we have two different situations.
\begin{figure}[t]
\includegraphics[angle=0.0, width=7.5cm]{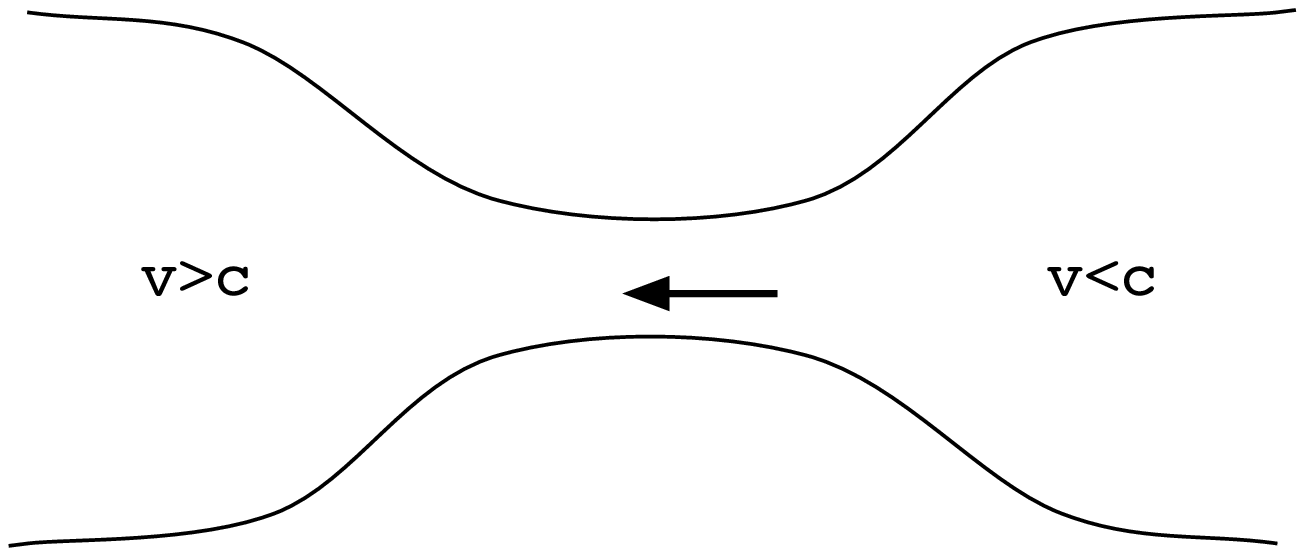}\hfil
\includegraphics[angle=0.0, width=7.5cm]{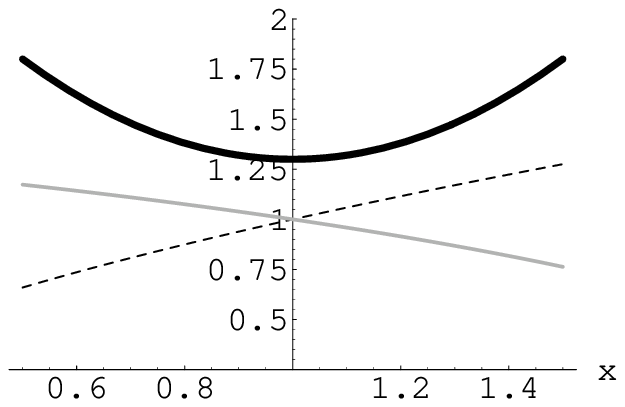}
\caption{\label{frame1}
Flow situation (on the left) and qualitative behavior of the 
parameters (on the right) for a converging-diverging  nozzle. 
The black, grey and dashed lines   represent  the tube cross-section $A$,  
 the fluid velocity $v$ and the
speed of sound $c$, respectively.
 The vertical 
axis indicates the position of the horizon. $c$ and $v$ are 
normalized to their horizon values.} 
\end{figure}

\begin{figure}[b]
\includegraphics[angle=0.0, width=7.5cm]{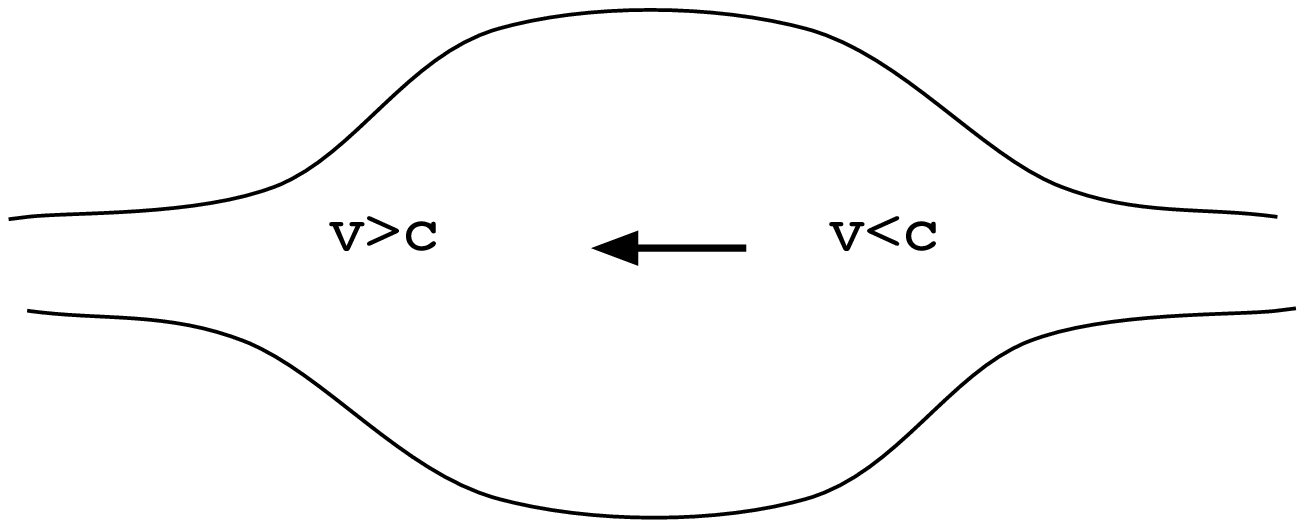}\hfil
\includegraphics[angle=0.0, width=7.5cm]{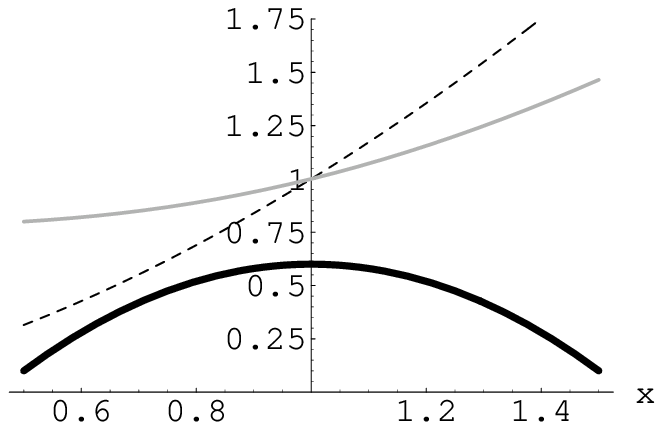}
\caption{\label{frame2}
Flow situation (on the left) and qualitative behavior of the 
parameters (on the right) for a  diverging-converging  nozzle. 
The black, grey and dashed lines   represent  the tube cross-section $A$,  
 the fluid velocity $v$ and the
speed of sound $c$, respectively.
 The vertical 
axis indicates the position of the horizon. $c$ and $v$ are 
normalized to their horizon values.} 
\end{figure}

{\it(i)} The fluid velocity grows monotonically passing from the
subsonic to the supersonic region. In this case the cross-section $A$ has a local
{\it{minimum}} at the horizon. This corresponds to the usual situation.
The flux tube has the shape of the Laval
nozzle, a converging pipe where the fluid is accelerated, followed by a
throat  and by a diverging pipe where the fluid continues to accelerate.
The corresponding flow situation  and 
qualitative behavior  of the parameters are  shown in Fig. 
(\ref{frame1}).

{\it (ii)} The fluid velocity decreases monotonically passing from the
subsonic to the supersonic region. In this case the cross-section $A$
has a local {\it maximum} at the sonic horizon.  This flow situation
is  very unusual and until now has not be considered in the
literature.  The flux tube has the form of a diverging pipe where
the fluid is decelerated followed  by a mouth  and by a converging pipe where
the fluid continues the deceleration.  The corresponding flow situation  and 
qualitative behavior  of the parameters are  shown in Fig. 
(\ref{frame2}).

Notice that if the tube cross-section is given as a  function of the
fluid velocity,
we can write the nozzle equation in the equivalent form
\beq\lb{nozzle1}
\frac{dA}{dv}=\frac{A}{v}(M^{2}-1).
\feq
This equation relates the appearance of an horizon directly with the
existence of local extrema of $A(v)$ without the need of further
conditions on the behavior of the space derivatives of $v$.
We will make use this equation when discussing the formation of
horizons in a fluid with spherically symmetric flow.

Until now our discussion has been focused on the nozzle equation
(\ref{nozzle}). The previous results are the whole information we can
extract  from this equation. There are two main points on which Eqs.
(\ref{nozzle}) give no information. First, they do not say to us
which of the two possibilities ({\it{i}}), ({\it{ii}}) is actually realized.
Second, our results are based on the rather strong assumption
$v'(x)\neq\infty,0$. In general its  validity depends on both the
dynamics and the initial conditions \cite{Cardoso:2004fi}.
In order to answer to the previous questions one has to choose an
equation of state for the fluid and to solve   Eqs.
(\ref{de}). To be more concrete we will consider a fluid with a
power-law equation of state,
\beq\lb{es}
P=\frac{a^2}{n} \rho^{n}.
\feq
where $a,n\neq 0$ are  real  constants.
This equation of state describes almost all physically  interesting
fluids: perfect fluid ($n=1$), Bose-Einstein condensate ($n=2$),
Chaplygin gas ($n=-1$).
Even in the simple one-dimensional case under consideration
explicit solution of Eqs. (\ref{de}),(\ref{es})  cannot be found.
However, one can find solutions in implicit form, writing $\rho,c,A$
as a function of the
fluid velocity $v$. The solution of Eqs. (\ref{de}) for a fluid with
equation of state given by  Eq. (\ref{es}) and $n\neq 1$ reads,

\beq\lb{sol}
\rho(v)=\biggl(\frac{n-1}{2a^{2}}(\alpha-v^{2})¥\biggr)^{\frac{1}{n-1}},\quad
c^{2}(v)= \frac{n-1}{2}(\alpha-v^{2}),\quad
A(v)= \frac{\beta}{\rho v},
\feq
where $\alpha, \beta $ are integration constants determined by the
initial conditions.
Formation of the sonic horizon requires $c(v_{h})=v_{h}$, which
yields for $n\neq -1$
\beq\lb{bc}
v_{h}=\sqrt{\frac{n-1}{n+1}\alpha},
\feq
where $v_{h}$ is the fluid velocity at the horizon.
Formation of the sonic horizon is possible in the following
range of variation of the  parameters.
For $n>1$,  $\alpha>0$ and $v<\sqrt\alpha=\sqrt{(n-1)/(n+1)}\,v_{h}$.
For $-1 <n<1$,  $\alpha<0$. For $n<-1$, $\alpha>0$, $v>\sqrt{\alpha}$.
The Chaplygin gas ($n=-1$) represents a limiting case. Formation of
the horizon requires  $\alpha=0$, which in turns implies $A=const$
and $v=c$, identically.

Using Eqs. (\ref{sol}) we can also determine which of the two
possibilities {\it (i)}, {\it (ii)} is actually realized.
Differentiating two times $A(v)$ we get $(dA/dv)|_{v_{h}¥}=0$
and $(d^{2}A/dv^{2})|_{v_{h}}> 0$ ($< 0$) for $n> -1$ ($n<
-1)$. Thus, the horizon forms in correspondence of a local minimum
(maximum) of the function $A(v)$ for $n>-1$ (for $n<-1$). Moreover
$d^{2}A/dv^{2}$ is everywhere nonzero excluding the
possibility of a flex point. To infer about the behavior of $A(x)$,
once the behavior of $A(v)$ is known, we need just to use the
trivial identities $dA/dx=(dA/dv)(dv/dx),\,
d^{2}¥A/dx^{2}¥=(d^{2}¥A/dv^{2}¥)(dv/dx)^{2}+(dA/dv)(d^{2}v¥/dx^{2})¥$.
When $dv/dx$ is finite and nonvanishing, local maxima (minima) of
$A(v)$ correspond to local maxima (minima) of $A(x)$. We have
therefore shown that in the case of a fluid  with a  power-law
equation of state (\ref{es}), the acoustic horizon forms in
correspondence of a minimum (maximum) of the cross-section when
$n>-1$ ($n<-1$). The Chaplygin gas ($n=-1$) represents a limiting
case for which  the cross-section must  be constant and $v=c$ everywhere
along the tube.
\vskip 3truemm
\leftline{\it{Flux tube with  constant cross-section and
non-homogeneous external potential}}
\vskip 3truemm
This case can treated similarly to the previous case.
Setting
$A=const$  in Eqs. (\ref{de}) one derives an equation similar to the
nozzle equation (\ref{nozzle}), in which $A$ is traded for  $\psi$
\beq\lb{nozzle2}
\frac{d\psi}{dx}= -\frac{v}{M^{2}¥}\frac{dv}{dx}(M^{2}-1).
\feq
The main difference between Eq. (\ref{nozzle}) and (\ref{nozzle2}) is
the change of sign in the right hand side.  This means that,
differently  to what happens for $v'(x)$ and $A'(x)$ in the nozzle
Eq. (\ref{nozzle}),
$v'(x)$ and $\psi'(x)$ have the same (opposite) sign in
the subsonic (supersonic) region.

\begin{figure}[t]
\includegraphics[angle=0.0, width=7.5cm]{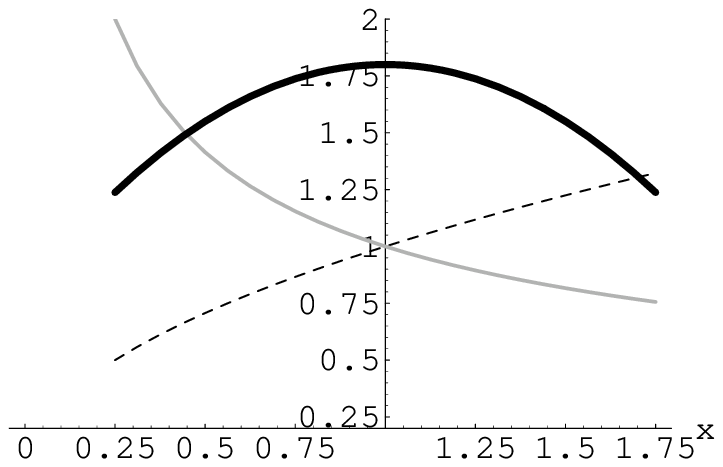}\hfil
\includegraphics[angle=0.0, width=7.5cm]{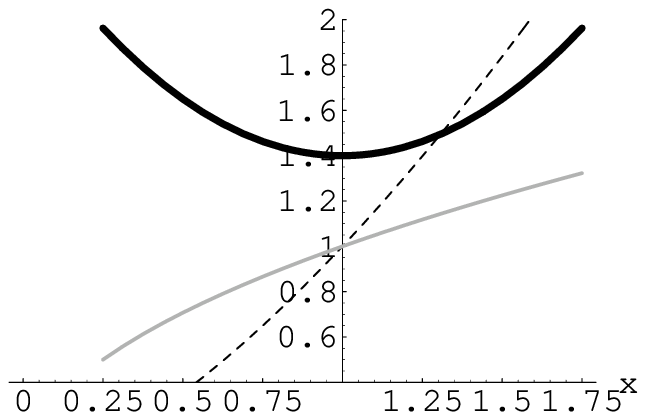}
\caption{\label{frame3}
Qualitative behavior of the 
parameters for a flux tube of constant section and varying external 
potential. The figure on the left (right) corresponds to case  
I, (II), respectively.
The black, grey and dashed lines   represent  the external potential  
$\psi$,  
 the fluid velocity $v$ and the
speed of sound $c$, respectively.
 The vertical 
axis indicates the position of the horizon. $c$ and $v$ are 
normalized to their horizon values.} 
\end{figure}

Again, assuming that $v$ and $v'(x)$ are everywhere finite and nonvanishing,
one can easily show using Eq. (\ref{nozzle2}) that the  horizon
must form   in correspondence of 
local extrema of the external potential $\psi$. This gives a a
null (external) force condition at the horizon location.
Depending on the sign of $v'(x)$ we have  two different situations.
{(I)} The fluid velocity grows monotonically passing from the
subsonic to the supersonic region. In this case the  external
potential $\psi$ has a local
{\it{maximum}} at the horizon.  
{(II)} The fluid velocity decreases monotonically passing from the
subsonic to the supersonic region. Now the external potential
has a local {\it minimum} at the sonic horizon.  
The qualitative behavior  of the parameters for both cases {(I)} 
and {(II)} is depicted in Fig. (\ref{frame3}).

Also in the case under consideration we can solve the fluid-dynamical
equations  for a fluid with equation of state  given by Eq.
(\ref{es}). We have for $n\neq 1$ 
\beq\lb{sol1}
\rho(v)= \frac{\beta}{v}, \quad c^{2}(v)= a^{2}\left(\frac{\beta}
{v}\right)^{n-1}, \quad \psi(v)=\frac{a^{2} }{1-n}\left(\frac{\beta}
{v}\right)^{n-1}-
\frac{v^{2}}{2} +{\alpha}, \feq where $\alpha, \beta $ are
integration constants. Solving the equation $c(v)=v$ one easily finds
the fluid velocity on the horizon,
$v_{h}=(a^{2}\beta^{n-1})^{1/(n+1)}$. Differently from the
case of a flux tube with nonconstant section,  in this case the
formation of the horizon does not imply limitations on the range of
variation of $v$.  Using the previous equations one can now easily
show that $(d\psi/dv)|_{v_{h}}=0$ and
$(d^{2}¥\psi/dv^{2}¥)|_{v_{h}¥}=-(n+1)$ so that
$(d^{2}¥\psi/dv^{2})|_{v_{h}¥}> 0$ ($<0$) for $n> -1$ ($n<
-1$). Using the same arguments used for the flux tube with 
non-constant section, one  concludes that for $n>-1$ ($n<-1$) the horizon
forms in correspondence  of a maximum (minimum) of the external
potential. The value $n=-1$ is also here a limiting case with the
same behavior as that previously discussed.

In solving Eqs. (\ref{de}) in the two cases of constant and
non constant cross-section, we have not considered  $n=1$  in the
equation  of state (\ref{es}). Solutions (\ref{sol}) and
(\ref{sol1})  become singular for  $n=1$ and this case has to be
considered separately.  For $n=1$ the solutions for a flux tube with
variable cross-section read
 \beq\lb{sol2} \rho(v)=\alpha
e^{-\frac{v^{2}}{2a^{2}}},\quad A(v)=\frac{\beta}{\alpha v}
e^{\frac{v^{2}}{2a^{2}}},\quad c=a.
\feq For a flux tube of
constant section  we have instead, \beq\lb{sol3}
\rho(v)=\frac{\beta}{v},\quad \psi(v)=-a^{2}\ln\frac{\alpha}{v}-
\frac{v^{2}}{2}, \quad c=a. \feq In both cases  we have for the
fluid velocity at the horizon $v_{h}=a$. A straightforward
calculation shows that, as expected, the horizon forms for a minimum
of the tube cross-section  and for a maximum of the external
potential.

As we have already noted    one main physical requirement for the
formation of an acoustic horizon is that $v'(x)$ remain  finite and
non vanishing throughout the flux tube. In general  to keep
$v'(x)\neq 0,\infty $ one needs a fine-tuning of   the initial
conditions. For instance, if $v$ is too large at the entrance of the
Laval nozzle the flow will become supersonic at a point  upstream from the
throat so that  $v'(x)\to -\infty$  \cite{Cardoso:2005ij}. This will result in
the generation of a shock wave, which will destroy  the stationary
flow of the fluid. We will not discuss here how this fine-tuning can
be realized. We will just point out  that the condition $v'(x)\neq 0,
\infty$ can be obtained as necessary consequence of imposing a
dynamical analogy between the fluid flow and a gravitational,
spherically symmetric black hole \cite{Cadoni:2004my,Cadoni:2005jg}. 

The Einstein equations
for a spherically symmetric black hole can be put in correspondence
with the fluid motion described by Eqs. (\ref{de})  and constrained
by 
\beq\lb{constraint} 2\frac{dY}{dr}-X\frac{dF}{dr}+
2e^{-F}\frac{d\psi}{dr}=\la^{2} V. \feq
where
$X=\frac{\rho}{c}(c^{2}-v^{2}),\, Y=\rho c, \,F=\ln\frac{c}{\rho},$
$r$ is the radial coordinate for the black hole spacetime, 
$\la^{2}¥$ is the inverse of Newton constant and $V$ is a function of
the coordinate $r$, which depends on the particular gravitational
model under consideration \cite{Cadoni:2005jg}. The constraint
(\ref{constraint}) represent a  constraint on the
geometrical form of the flux tube, which forces the cross-section to
have a local extremum exactly at the position of the horizon. The
constraint (\ref{constraint}) also  implies that $v'(x)$ is finite
and nonvanishing at the horizon. This can be demonstrated  using the
trivial identity
$dv/dx=(dv/dY)(dY/dX)(dX/dr)(dr/dx)$
and showing that each factor in the product is individually finite
and nonvanishing at the horizon.  The third factor is proportional
to the Hawking temperature of the horizon, the fourth is the fluid
density \cite{Cadoni:2004my,Cadoni:2005jg},  therefore  both must be 
$\neq 0, \infty$.  The
first and the second can be shown to be  finite and non-vanishing at
the horizon by using the explicit solution of the constrained fluid
dynamics derived in Ref. \cite{Cadoni:2005jg}.

\vskip 3truemm
\leftline{\bf Spherically symmetric flow}
\vskip 3truemm
Let us now consider the inviscid, isentropic, spherically symmetric
flow of a fluid with equation of state given by (\ref{es}). For
simplicity we will set to zero the external forces acting on the
fluid. Using spherical coordinates $(r,\theta,\phi)$, the continuity
and Euler equations yield, 
\beq\lb{ss} \frac{d(r^{2}
v\rho)}{dr}=0,\quad\quad
v\frac{dv}{dr}+a^{2}\rho^{n-2}\frac{d\rho}{dr}=0. \feq 
These
equations can be easily generalized to the case when a pointlike
source is located at the origin of the spherical coordinate system.
Whereas the Euler equation remains unchanged the continuity equation
gets a contribution proportional to $\delta(r)$. In this more
general case for $r>0$ the dynamics of the system is still described by
Eqs (\ref{ss}). 

The formation of sonic horizons for the spherically
symmetric flow can be easily discussed noticing that Eqs. (\ref{ss})
are formally identical to those describing an axi-symmetric flow
along a tube with cross-section $A(r)= 4\pi r^{2}$. Because the
section increases monotonically, in view of the previous derived
results we should conclude that  an horizon cannot form. However,
these results rely on the assumption that $v'(r)=dv/dr$ is
everywhere finite. Relaxing this assumption could allow for the
formation  of an acoustic horizon even though $A(r)$ does not have
 local extrema. Divergence of $v'(r)$ implies that the fluid has
infinite acceleration. 
The discussion  has therefore no direct
physical relevance. However, considering this more general situation
sheds light on the physical mechanism on which  horizon
formation is based. 

Setting $\beta=r^{2}v \rho$,  Eqs.  (\ref{ss})
are solved by Eqs. (\ref{sol}) with $A(r)= 4\pi r^{2}$.  The
condition for horizon formation is also given by Eq. (\ref{bc}).
Using  Eqs. (\ref{sol}) and the continuity equation in (\ref{ss})
one  gets $A(v)=4\pi
\frac{\beta}{v}[\frac{n-1}{2a^{2}}(\alpha-v^{2})]^{1/(n-1)}$. One
can easily show that for $n>-1$ ($n<-1$) $A(v)$   has a minimum
(maximum) in correspondence of  $v=v_{h}$, with $v_{h}$ given by
Eq. (\ref{bc}). It is immediately evident from the equation
$dA/dv=(dA/dr)(dr/dv)$ that in this case 
local extrema of the function $A(v)$ are not related with local
extrema of $A(r)$ but with the divergence of $dv/dr$.
The function $A(v)$ contains more information about the
formation of acoustic horizons then the function $A(x)$ (or $A(r)$).
The vanishing of $dA/dv$ is a necessary and sufficient condition 
for the existence of a point where $v=c$,  .


\end{document}